\documentclass{svproc}
\usepackage{url}
\usepackage{amsmath}
\usepackage{bbm}
\usepackage{comment}

\begin{document}
\mainmatter              
\title{A multiscale Bayesian nonparametric framework for 
partial hierarchical clustering}
\titlerunning{Multiscale model for partial hierarchical clustering}  
%
\author{Lorenzo Schiavon\inst{1} \and Mattia Stival\inst{1}
}
\authorrunning{Lorenzo Schiavon \& Mattia Stival} 
%
\tocauthor{Lorenzo Schiavon, Mattia Stival}
\institute{Ca’ Foscari University of Venice, Department of Economics, Italy.\\
Corresponding address: \email{lorenzo.schiavon@unive.it}} 

\maketitle              

\begin{abstract}
In recent years, there has been a growing demand to discern clusters of subjects in datasets characterized by a large set of features. Often, these clusters may be highly variable in size and present partial hierarchical structures.
In this context, model-based clustering approaches with nonparametric priors are gaining attention in the literature due to their flexibility and adaptability to new data. However, current approaches still face challenges in recognizing hierarchical cluster structures and in managing tiny clusters or singletons.
To address these limitations, we propose a novel infinite mixture model with kernels organized within a multiscale structure. 
Leveraging a careful specification of the kernel parameters, our method allows the inclusion of additional information guiding possible hierarchies among clusters while maintaining flexibility. We provide theoretical support and an elegant, parsimonious formulation based on infinite factorization that allows efficient inference via Gibbs sampler. 
\end{abstract}

\section{Introduction}
Clustering applications in high-dimensional settings are spreading in last years.
Customer base segmentation
, topic modeling
, and grouping of survey respondents
arise common challenges due to the presence of highly variable cluster sizes and partial hierarchies. For example, if we are interested in clustering European city districts based on age-friendliness, overlapping hierarchical structures may emerge, reflecting nested administrative regions, as well as district similarities, and other characteristics as the proximity to city centers. In other words, datasets often exhibit non-exchangeable partitions due to dependence structures among observations.

To group $n$ subjects, Bayesian literature often employs model-based clustering, assuming a multimodal density function for the $p$-variate data vector $y_i$ of the subject $i$ expressed as:
$$
p(y_i) = \sum_{h=1}^{k} \pi_{h} \mathcal{K}_h(y_i),
$$
where $\pi_{h};(h=1,\ldots,k)$ is a suitable sequence of probabilities such that $\sum_{h=1}^{k} \pi_h= 1$, and $\mathcal{K}_h(y_i);(h=1,\ldots,k)$ are the density functions, or kernels.
In such a construction subjects may be grouped according to their corresponding kernel in a data-augmentation representation.
Bayesian nonparametric methods \cite{muller2015} extend this to infinite mixture models ($k=\infty$), dynamically adapting the number of clusters to the data, providing a balanced approach to model fit and complexity. 
The core assumption involves random variables drawn from an unknown probability distribution, itself drawn from a prior distribution. 
While the Dirichlet process and Pitman-Yor process 
are commonly used priors, they exhibit a ``rich-get-richer'' property
, influencing partitions with a few large clusters.
However, they also tend to produce tiny clusters or singletons and struggle to capture hierarchical cluster structures due to their exchangeable nature and the independent distributed prior on kernels. 
Recognizing the correct number of clusters becomes challenging, especially in high-dimensional settings, as demonstrated by
 \cite{chandra2023}.

In response to these challenges, we propose a novel infinite multiscale  mixture model based on the structure:
\begin{equation}
\label{eq:multiscale}
    p(y_i) = \sum_{s=0}^{\infty}\sum_{h=1}^{2^s} \pi_{s,h} \mathcal{K}_{s,h}(y_i),
\end{equation}
where there are $2^s$ kernels for every scale level $s$, and an increase in the level $s$ indicates a decrease in the scales. 
Unlike existing multiscale mixture models \cite{canale2016}, \cite{stefanucci2021}, our method organizes not only weights but also kernels within a multiscale structure. 
This induces a hierarchical structure on partitions similar to the nested Chinese restaurant process prior proposed by \cite{blei2010}, a hierarchical extension of the Dirichlet process prior capable of generating tree-structured clusterings. To overcome the limitations of the latter approach, such as a fixed tree depth, lack of control over the branching factor, and the inability to accommodate partial hierarchies, we enhance the model flexibility and account for additional information.
Our prior incorporates covariates in branch selection of the multiscale structure, similar to the approach of \cite{horiguchi2022}, enabling the generation of flexible tree-structured clusterings characterized by partially overlapping partitions covariate-related that well adapt to the multiscale and possible hierarchical nature of the data.

A convenient formulation expresses our model in terms of a factorization of the data matrix into a kernel-parameter matrix and a suitable selection matrix. This allows us to leverage recent advancements in the infinite factorization literature to derive efficient inference via Gibbs sampling.

The paper is organized as follows: Section 2 presents the parameter structure for our proposed multiscale infinite mixture model. The formulation based on infinite factorization is introduced in Section 3. Section 4 reports the inference algorithm, while Section 5 concludes mentioning some possible applications and final remarks.

\section{Multiscale infinite mixture model}
We consider the mixture infinite multiscale model defined in equation \eqref{eq:multiscale}.
It may be convenient considering kernel functions belonging to the same distribution family. In particular, location-scale family may represent an appealing solution able to balance flexibility and dimensionality reduction. Under this assumption, we can rewrite $\mathcal{K}_{s,h}(y_i)$ as $\mathcal{K}(y_i;\mu_{s,h}, \sigma_{s,h})$, where $\mu_{s,h}$ and $\sigma_{s,h}$ are the kernel-specific location and scale parameters, respectively.

A prior measure for the multiscale mixture is obtained by specifying a stochastic process for the infinite vector $\pi$ and the parameter tensors $M$ and $\Sigma$ of the kernels.

\subsection{Weight parameters}
We first focus on the stochastic process that models the vector $\pi$. We introduce a sequence of independent random variables $w_s;(s=0,\ldots,\infty)$ with $w_0=1$, and $w_s$ taking values in $(0,1)$ for $s>0$. The term $w_s$ describes the probability of taking the right path from scale $s$ to scale $s+1$, conditioned on not stopping at the node $(s,h)$. In the literature on multiscale mixture models \cite{canale2016, stefanucci2021}, it is common to also introduce a further sequence of random variables $u_s;(s=0,\ldots,\infty)$ modeling the probability of stopping at level $s$.
Thus, the weights are defined as
\begin{equation}
\label{eq:weights1}
 \pi_{s,h} =  u_s\prod_{t\leq s} \tilde{w}_{sht}   \quad \text{with} \quad
 \tilde{w}_{sht} = 
 \begin{cases}
     w_t \quad \text{if}  \; \lceil h/2^{(s-t)} \rceil \; \text{is even or } s=0\\
     1- w_t  \quad \text{if}  \; \lceil h/2^{(s-t)} \rceil \; \text{is odd and } s\neq 0.
 \end{cases}
\end{equation}
The stopping probability $u_s$ is specified through a stick-breaking construction
$u_s = \tilde{u}_s \prod_{q>s} (1-\tilde{u}_q)$, where $\tilde{u}_s$ is the probability of being in level $s$, conditioned on not being in level $s+1$.
For the sake of parsimony in the number of parameters, we set $\tilde{u}_s = w_{s+1}$, such that
\begin{equation}
\label{eq:weights2}
 \pi_{s,h} =  w_{s+1} \bigg(\prod_{q>s+1} (1-w_{q}) \bigg) \prod_{t\leq s} \tilde{w}_{sht}.
\end{equation}
Contrary to the literature \cite{stefanucci2021},  it is not necessary for the weights to decrease in expectation over the index $s$. In fact, the following theorem states a sufficient condition to ensure an increasing expectation of the prior total weight $\pi_s = \sum_{h\leq 2^s} \pi_{s,h}$ over the index $s$ for $s>0$.
\begin{theorem}
Let the weights $\pi_{s,h}\;(s=1,2,\ldots;\,h=1,\ldots,2^s)$ be defined as in \eqref{eq:weights1}--\eqref{eq:weights2}.
If $E(w_{s+1})\geq E(w_{s})>0$, then
$E(\pi_{s+1})>E(\pi_s)$.
\end{theorem}
Similarly, we allow for specifying the weight process such that the variance of the weights is not necessarily decreasing according to the level of the tree.
In fact, this decreasing behavior typically characterizes the weights generated via stick-breaking construction, making the uncertainty quantification unreliable  (see \cite{horiguchi2022} for a discussion about this topic).
Flexibility in choosing the $w_s$ prior and, consequently, on the behavior of the weights process should not prevent the model from guaranteeing that $\text{pr}(M, \Sigma)$ is a proper random mixture measure, with $\delta_x$ indicating the Dirac delta function, such that $f(y) = \int K(y; M, \Sigma) d\text{pr}(M,\Sigma)$.
Given the specification in \eqref{eq:weights1}--\eqref{eq:weights2}, it is straightforward to verify that 
\begin{equation*}
    \sum_{s\geq 0} \sum_{h=1}^{2^s} \pi_{s,h} =  1-\prod_{q>0}  (1-w_{q}).
\end{equation*}
In view of this, we define a further kernel  $\mathcal{K}(y_i;\mu_{\infty,1} \sigma_{\infty,1})$ with weight $\pi_{\infty,1} = \prod_{q>0} (1-w_{q})$ such that the weights sum to one, implying a proper definition of $\text{pr}(M,\Sigma)$ as a random mixture measure.

\subsection{Location parameters}
To ensure sufficient flexibility, especially in a high-dimensional context, one may choose to independently define the location parameters. This would imply that kernel locations are primarily influenced by the data, without the need to impose any structured partition on the sample space. However, adopting this approach could result in an explosion of model complexity, leading to computational challenges and overfitting behaviors. This is attributed to the exponential relationship between the scale level $s=1,\ldots,$ and the dimension of the location parameter induced by the tree structure. Generalizing to non-binary trees would only exacerbate these issues.

We propose a stochastic process for the location parameter such that the number of parameters grows linearly with the scale level of the tree. 
We define a $p\times k$ matrix $\Theta$, where the column $s+1\,(s=0,1,\ldots,)$ is a $p$-variate independent random vector $\theta_{s}$.
Then, the location parameter vector of the kernel $(s,h)$ is  
\begin{equation}
\label{eq:location}
    \mu_{s,h} = \theta_{s}+ \sum_{t< s} \theta_{t} \tau_{s,h,t}  \quad \text{with} \quad
 \tau_{s,h,t} = 
 \begin{cases}
     1 \quad \text{if}  \; \lceil h/2^{(s-t)} \rceil \; \text{is even or } s=0\\
     0  \quad \text{if}  \; \lceil h/2^{(s-t)} \rceil \; \text{is odd and } s\neq 0.
 \end{cases}
\end{equation}
for $s\geq 0$ and $\mu_{\infty,1}=0$.
We specify an increasing shrinkage prior over $\Theta$, defining a Gaussian continuous mixture on the entry of the matrix  
\begin{equation}
\label{eq:theta-prior}
    \theta_{sj} \sim N(0, \phi_{sj}^2 \gamma_s ),
\end{equation}
where $\phi_{s}^2=(\phi_{s1}^2,\ldots, \phi_{sp}^2)^\top \; (s=0,1,\ldots)$ are nonnegative independent and identically distributed local scale vectors, while $\gamma_s\; (s=0,1,\ldots)$ are nonnegative independent column scales characterized by a decrease in expectation, i.e. $E(\gamma_s)>E(\gamma_{s+1})\;(s=0,1,\ldots)$.
This specification induces a stochastically decreasing behaviour over $\theta_s$ such that the kernels in deeper levels of the tree are located increasingly closer.

Such behavior aids the inference procedure of the model. Indeed, infinite mixture models can be challenging to handle in practice due to their infinite nature. It is common in the literature to induce a stochastically decreasing behavior over the index $s=1,2,\ldots,$ of the weights $\pi_{s,h}$ toward zero, aiming to approximate the model with a truncated version of the mixture by setting all the elements $\pi_{s,h}$ to zero for every $s$ greater than a certain large value $L$. In our case, the weights are not necessarily decreasing, and therefore, we exploit the process defined on the kernel parameters to approximate the model through a finite mixture model. For every $s>0$, we collapse the kernel $\mathcal{K}(\mu_{s,h}, \sigma_{s,h})$ onto a kernel of the higher level of the tree $\mathcal{K}(\mu_{s-1,l}, \sigma_{s-1,l})$ if the prior probability that the parameters of the two kernels are close is sufficiently high.
In practical applications, the hierarchical construction of cluster means, as explained above, exhibits several similarities with agglomerative hierarchical clustering (see \cite{nielsen2016} for a recent review), where $\zeta$ functions as a threshold to cut the dendrogram tree, defining a set of clusters.

\subsection{Scale parameters}
To select the process for the scale parameters, we employ similar reasoning to that discussed for the location parameters. The main difference lies in the fact that $\sigma_{s,h}$ is no longer a $p$-variate vector but is a scale $p \times p$ matrix.
The idea is to define a process $\sigma_{s,h};(s=1,2,\ldots;,h=1,\ldots,2^s)$ with a decreasing behavior over $s$ allowing us to truncate out the kernels with lower parameter scales in the inference procedure.
We specify the scale of the largest kernel $\mathcal{K}(y_i; \mu_{\infty,1},\sigma_{\infty,1})$ as
$$
\sigma_{\infty,1} = \sum_{q\geq 0} (\tilde{\lambda}_{q} \tilde{\lambda}_{q}^\top +\lambda_{q} \lambda_{q}^\top)+\varSigma,
$$
with diagonal matrix $\varSigma=\textnormal{diag}(\varsigma_{1},\ldots,\varsigma_{p})$, and $\lambda_s$ and $\tilde{\lambda}_s$ the column vectors of the left path matrix $\Lambda$ and the right path matrix $\tilde{\Lambda}$, respectively. 
Since we are interested in a decreasing scale behavior over the deep level of the tree, we specify the other scales $\sigma_{s,h}$ for $s\geq 0$ and $h=1,\ldots,2^s$ through successive subtractions
\begin{align}
\label{eq:scale}
    \sigma_{s,h} &= \sigma_{\infty,1} - \sum_{t<s} \tilde{\lambda}_{t} \tilde{\lambda}_{t}^\top (1-\tau_{s,h,t}) - \sum_{t<s} \lambda_{t} \lambda_{t}^\top \tau_{s,h,t} -\lambda_s \lambda_{s}^\top, \\
    \tau_{s,h,t} &= 
 \begin{cases}
     1 \quad \text{if}  \; \lceil h/2^{(s-t)} \rceil \; \text{is even or } s=0\\ \nonumber
     0  \quad \text{if}  \; \lceil h/2^{(s-t)} \rceil \; \text{is odd and } s\neq 0.  \nonumber
 \end{cases}
\end{align}
On generic elements $s,j$ of $\Lambda$ and $\tilde{\Lambda}$, we put the prior 
\begin{equation}
\label{eq:lambda-prior}
    \lambda_{sj} \sim N(0, \psi_{sj}^2 \gamma_s), \quad \tilde{\lambda}_{sj} \sim N(0, \tilde{\psi}_{sj}^2 \gamma_s),
\end{equation}
where $\psi_{s}^2=(\psi_{s1}^2,\ldots, \psi_{sp}^2)^\top$ and $\tilde{\psi}^2_{s}=(\tilde{\psi}^2_{s1},\ldots, \tilde{\psi}^2_{sp})^\top\; (s=0,1,\ldots)$ are nonnegative independent and identically distributed local scale vectors, while $\gamma_s\; (s=0,1,\ldots)$ is the tree level-specific scale defined on the location process. 

The rationale behind this specific construction is to have scale matrices that exhibit the following three properties: an additive structure such that the parameter dimension grows linearly with the level of the tree, a unique characterization for each kernel within the tree structure, and a decreasing magnitude behavior of the scale as we delve deeper into the levels of the tree structure.
Related to the latter point, if we approximate the scale magnitude of a certain kernel with the trace of the scale matrix
, it is straightforward to demonstrate that for each $h=1,\ldots,2^{s+1}$, there always exists an $l=1,\ldots,2^s$ such that:
$\textnormal{tr}(\sigma_{s+1,h}) = \textnormal{tr}(\sigma_{s,l}) - \textnormal{tr}(\lambda_{s+1}\lambda_{s+1}^\top)$, with $\textnormal{tr}(\lambda_{s+1}\lambda_{s+1}^\top)$ a  positive constant by construction.

In a location-scale family with two parameters, one can define two similar kernels by specifying similar location and scale parameters.
Thus, considering the spectral norm as a two-dimension extension of the euclidean norm in two dimensions we may provide the following definition.
\begin{definition}
\label{def:collapse}
Consider the set of kernels $\mathcal{K}(\mu_{s,h}, \sigma_{s,h})\;(s=0,1,\ldots;\,h=1,\ldots,2^s)$ with distributions belonging to a location-scale family with all parameters set equal but for the location and scale parameters.
We say that a kernel $\mathcal{K}(\mu_{s,h}, \sigma_{s,h})$ is collapsing a priori on a kernel $\mathcal{K}(\mu_{q,l}, \sigma_{q,l})$, with $q<s$, if 
\begin{equation}
\label{eq:collapse}
\textnormal{pr}(||\mu_{s,h}-\mu_{q,l}||_2<\zeta_\mu)> \xi_\mu, \quad \textnormal{and} \quad \textnormal{pr}\{||\sigma_{q,l}-\sigma_{s,h}||_2<\zeta_\sigma\}> \xi_\sigma,
\end{equation}
for certain constants $\zeta_\mu, \zeta_\sigma>0$ and $\xi_\mu, \xi_\sigma \in (0,1)$.
\end{definition}
Collapsing kernels can be neglected in practice. If the probability \eqref{eq:collapse} is increasing over the index $s$, one can truncate the model for $s>L$ with $L$ a large value.
The following theorem introduce sufficient conditions to guarantee that kernels collapse a priori for every sufficiently deep tree level. 
to Theorem \ref{th:loc-collapse} to guarantee that kernels collapse a priori for every sufficiently deep tree level. 
\begin{theorem}
     Given the location parameter process \eqref{eq:location}, the scale matrix process \eqref{eq:scale}, and the prior specification \eqref{eq:theta-prior}, \eqref{eq:lambda-prior}. If the expected values of $\gamma_s\;(s=1,2,\ldots)$ are a convergent series, then there exists a level $L \in \mathbbm{N}$ such that all the kernels $\mathcal{K}(\mu_{s,h},\sigma_{s,h})$ for every $h=1,\ldots,2^s$ and every $s>L$ are collapsing kernels on kernels $\mathcal{K}(\mu_{q,l},\sigma_{q,l})$ with $q\leq L$, under definition \ref{def:collapse}.
\end{theorem}
This construction allows for the construction of a multiscale mixture model where the weights of the kernels are not forced decreasing along with the scales and such that the dimension of kernel parameter grows linearly with the level of the tree structure.

\section{Factorization clustering}
The model described in Section 2 can be rephrased to facilitate the development of an efficient computational procedure. We employ a parameter expansion construction, defining the $n$ latent $k$-variate binary vectors $\rho_i$ such that the element $\rho_{i(s+1)}$ is equal to $1$ with a probability of $w_{s}$. 
Under such expansion, the data vactor $y_i$ can be modelled as
\begin{equation}
\label{eq:factor-cluster}
    y_i = \Theta \rho_i + \epsilon_i \quad \epsilon_i \sim \mathcal{K}( 0, \tilde{\Lambda} \tilde{\rho}_i \tilde{\rho}_i^\top \tilde{\Lambda}^\top + \Lambda(\mathbbm{1}_k-\rho_i)(\mathbbm{1}_k-\rho_i)^\top \Lambda^\top +\varSigma),
\end{equation}
where $\tilde{\rho}_i$ is such that the element $s$ is specified as $\tilde{\rho}_{is} = \rho_{is} + \prod_{t \geq s} (1-\rho_{it})$.
This expanded representation arises from the intuitive depiction of a mixture model as a clustering model. In this framework, the subject $i$ is assigned, with a probability $\pi_{s,h}$, to the group $(s,h)$ characterized by a distribution density function $\mathcal{K}(y; \mu_{s,h}, \sigma_{s,h})$. 

\subsection{Gaussian multiscale mixture model and priors}

In case of Gaussian kernels we can rewrite the model in \eqref{eq:factor-cluster} as 
\begin{equation}
\label{eq:gauss-mod}
    y_i = \Theta \rho_i +  \tilde{\Lambda} \tilde{\eta}_i + \Lambda \eta_i + \epsilon_i, \quad \epsilon_i \sim N(0, \varSigma),
\end{equation}
with $\tilde{\eta}_i = \tilde{\rho}_i \tilde{z_i}$ and $\eta_i = (\mathbbm{1}_k-\rho_i) z_i$, where $\tilde{z}_i$ and $z_i$ are standard Gaussian random variables.

As mentioned earlier, we induce a decreasing behaviour over the column index $l$ of the matrices $\Theta$, $\Lambda$ and $\tilde{\Lambda}$. Let $l$ denote a column such that $s=\lfloor {l}/({c-1}) \rfloor$. We specify a continuous spike and slab prior on the level-specific parameter scale $\gamma_s$ 
\begin{equation*}
    \gamma_s \sim (1-\varpi_{s}) f_{\text{IG}}(a_\gamma, b_\gamma)+ \varpi_{s} f_{\text{IG}}(a_\gamma, \vartheta b_\gamma), 
\end{equation*}
with $0<\vartheta <1$ small enough, $f_{\text{IG}}(a_\gamma, b_\gamma)$ denoting the density function of an inverse gamma distribution with shape $a_\gamma$ and scale $b_\gamma$, and increasing spike probability 
\begin{equation*}
    \varpi_s = \sum_{t\leq s} \nu_t \prod_{m \leq t} (1-\nu_m), \quad 
\nu_m \sim \text{Beta}(1, a_\nu).
\end{equation*}
The hyperparameter $a_\nu$ is equal to the prior expected number of non collapsing level $\sum_{s=0}^\infty (1-\varpi_s) = a_\nu$, suggesting setting $a_\nu$ in accordance to the expected scale level of the problem.
Similarly to \cite{kowal2023}, conditionally to $\phi_{js}, \psi_{js}$, and $\psi_{js}$, the elements $\theta_{js}$, $\lambda_{js}$, and $\tilde{\lambda}_{js}$ are  distributed \textit{a priori} as normal mixture of inverse gamma distributions.

To accommodate for local behaviours, we define the local scale priors
\begin{equation*}
    \phi_{jl}, \psi_{jl}, \tilde{\psi}_{jl} \sim 0.5 f_N(1, 1) + 0.5 f_N(-1,1),
\end{equation*}
with $f_N(\mu,\sigma^2)$ indicating the density function of a normal distribution with mean $\mu$ and variance $\sigma^2$.
Under this construction, we can express the elements $\theta_{jl} = \phi_{jl} \theta_{js}^{*}$, $\lambda_{jl} = \psi_{jl} \lambda_{js}^{*}$, and $\tilde{\lambda} =  \tilde{\psi}_{jl} \tilde{\lambda}_{js}^{*}$ through the parameter expansion proposed by \cite{kowal2023} where the local parameters $\phi_{jl}$, $\psi_{jl}$, and $\tilde{\psi}_{jl}$ scatter the normal mixture inverse gamma distributed parameters $\theta_{jl}^*$, $\lambda_{jl}^*$, and $\tilde{\lambda}_{js}^{*}$, favouring the mixing of the posterior distribution sampling.

The subject-specific latent binary vectors $\rho_i$ are modelled by the prior
\begin{equation*}
    \rho_{is} \sim \text{Mult}_{c-1}(1, \pi_{is}), \quad
\pi_{is}\sim  \text{MultiProbit}({x}_i^\top B^{(s)} )
\end{equation*}
where $x_i^\top$ is a $d$-variate vector of covariates possibly informing on the kernel weights of the subject $i$, and $B^{(s)}$ is a $d \times (c-1)$ coefficient matrix with standard Gaussian independent prior on its elements $\text{vec}(B^{(s)}) \sim N(0, I_d \otimes I_{c-1})$, with $\otimes$ indicating the Kronecker product.

The prior eliciation is completed specifying $\Sigma =
\text{diag}(\sigma_1^2,\ldots, \sigma_p^2)$ with $\sigma_j^{-2} \sim
 \text{Ga}(a_\sigma; b_\sigma)$.

\section{Posterior inference}

Posterior inference can be conducted via Gibbs sampling.
Let $k$ denote both the level of depth at which the tree of clusters is truncated and the number of columns of $\Theta$, $\Lambda$, $\tilde{\Lambda}$.
Then, the generic iteration of the Gibbs sampler follows the steps below.
\begin{enumerate}
    \item Given $\Theta$, $\Lambda$, $\tilde{\Lambda}$, $\Sigma$, $z$,  $\tilde{z}$, and $B^{(s)}$, we update the $n$ vectors $\rho_i\,(i=1,\ldots,n)$ in parallel, by sampling from the Bernoulli full conditional distribution.
    The corresponding element of $\tilde{\rho}$ is updated by $\tilde{\rho}_{is} = \rho_{is} + \prod_{t \geq s} (1-\rho_{it})$.
    Since the posterior distribution depends on the entire vector $\rho_{i}$, the update is sequential with respect to the index $s$.

    \item  Given $\rho$, we update in parallel the $k$ coefficient vectors $B^{(s)}$, by exploiting the algorithm for the sampling of regression coefficients in probit models  based on a suitable representation of the unified skew-normal posterior for $B^{(s)}$ proposed by \cite{durante2019}.

    \item Given $\Theta$, $\Lambda$, $\tilde{\Lambda}$, $\rho$,  we update the $n$ latent observation $z_i$ and $\tilde{z}$, sampling from the corresponding Gaussian full conditional distribution.

    \item Given $\Theta$, $\Lambda$, $\tilde{\Lambda}$, $\phi$, $\psi$, $\tilde{\psi}$, we update the level-specific scale $\gamma_s$. Consider the decomposition $\theta_{jl} = \phi_{jl} \theta_{js}^{*}$, $\lambda_{jl} = \psi_{jl} \lambda_{js}^{*}$, and $\tilde{\lambda} =  \tilde{\psi}_{jl} \tilde{\lambda}_{js}^{*}$ such that $\theta_{js}^{*}$, $\lambda_{js}^{*}$, and $\tilde{\lambda}_{js}^{*}$ have zero mean normal mixture inverse gamma priors.
    Consistently with \cite{legramanti2020}, we define the independent indicators $\zeta_s$ assuming values $\{1,\ldots, k\}$ with prior $\text{pr}(\zeta_s = t) = \nu_t \prod_{m\leq t} (1-\nu_m)$. Then we sequentially update the augmented indicator $\zeta_s$ sampling from a multinomial distribution, the parameter $\nu_t$ from the beta full conditional distribution, and finally $\gamma_s^{-1}$ sampling from the proper gamma full conditional distribution.
    
    \item Given $\Lambda$, $\tilde{\Lambda}$, $\eta_i = (\mathbbm{1}_k-\rho_i)z_i$ and $\tilde{\eta} = \tilde{\rho}_i \tilde{z_i}$, we update the entries of $\Theta$,  
    following the data-augmentation strategy proposed in \cite{kowal2023}. 
    
    \item We follow an equivalent strategy to update the entries of  $\Lambda$ and $\tilde{\Lambda}$ matrices. 

    \item We finally update the error covariance matrix $\Sigma$ by sampling the opposite $\sigma_j^{-2}$ of the $p$ diagonal elements in parallel from the proper gamma full conditional distribution.
\end{enumerate}

The number of active levels is identified, at each iteration, as the number of levels $s$ such that $\zeta_s > s$. To speed up the computation one may implement an adaptive Gibbs sampler where the truncating level $k$ adapts with decreasing exponential probability as implemented by \cite{legramanti2020}.

Samples are aligned 
by reordering the levels, i.e. the order of the $\rho_i$ elements, to minimize the distance from a benchmark sample. 
Alternative efficient posterior approximation may be conducted by deriving a coordinate ascent optimization algorithm based on mean-field variational Bayes approximation.

\section{Possible applications and final remarks}

The model proposed in this paper is tailored to accommodate data sampled from multiscale mixture distributions, possibly characterized by nested modes. This framework corresponds to the data generating process in set of subjects organized in latent clusters presenting partial hierarchical group structures. Furthermore, potential hierarchies and interrelations among clusters can be informed by exogenous covariates, a common occurrence in various applications. For instance, in the clustering of European city districts based on age-friendliness, relevant information such as district location, socio-demographic composition, administrative region, and service presence can be included in our model as covariates $x$, inducing similar clustering paths $\rho_i$ and $\rho_l$ for comparable districts $i$ and $l$.

The additive structure promotes parsimony in the number of parameters, favoring the model's applicability in scenarios with high $p/n$ ratio. This is particularly relevant in genomics applications, where a vast set of gene expressions is measured on a small set of patients. A similar context is also common in topic modeling, where a moderate-to-small set of documents is clustered based on the frequency of a large set of words.

The convenient factorization facilitates efficient posterior approximation through Gibbs sampling and variational Bayes, making it adaptable to even large-scale problems. 

\subsection*{Funding} This work was developed within the project funded by Next Generation
EU - Age-It - Ageing well in an ageing society project (PE0000015), National Recovery and Resilience Plan (NRRP) - PE8 - Mission 4, C2, Intervention
1.3. The views and opinions expressed are only those of the authors and do not necessarily reflect those of the European Union or the European Commission. Neither the European Union nor the European Commission can
be held responsible for them.

%
%

\end{document}